\documentclass[12pt]{article}
\usepackage{graphicx}

\usepackage{epsfig}
\usepackage{latexsym}
\usepackage{amsmath}

\begin{document}

\title{   
Pitfalls in testing with linear  regression model by OLS
}

\author{  C.  Herteliu$^1$, B.V.  Ileanu$^1$, M. Ausloos$^{2,3,}$, G. Rotundo$^4$,
 \\ 
$^{1}$Department of Statistics and Econometrics, \\Bucharest University of Economic Studies,\\ Bucharest, Romania  \\Ê
Ê\\Ê$^{2}$School of Business, University of Leicester, \\ University Road, Leicester  LE1 7RH, UK  \\ \\ $^{3}$Group of Researchers for Applications of \\Physics to Economy and Sociology (GRAPES),\\ r. de la Belle Jardini\`ere 483, \\ B-4031 Liege, Wallonia-Brussels Federation
\\ \\ 
 \\$^{4}$ Sapienza University of Rome, Faculty of Economics,
\\Department of Methods and models for Economics, Territory and Finance,
\\via del Castro Laurenziano 9, I-00161 Roma, Italia \\
 }

 \date{\today}
\maketitle
 \vskip 0.5 cm

\begin{abstract}
This is a comment on Economic Letters 

DOI
http://dx.doi.org/10.1016/j.econlet.2015.10.015. We show that due to some methodological
aspects the main conclusions of the above mentioned paper should be a little bit altered.
 \end{abstract}

  \section{Introduction  }\label{sec:intro}
 Physical,  physiological and behavioral changes may occur in direct
response to environmental fluctuations that have an obvious and immediate
adaptive function.  Shimura et al.  (1981) were first to discuss in a modern way   "Geographical and secular changes in the seasonal distribution of births". Since then,  much data has been reported by seasonal effects time series. It seems that there is an 11-year cycle in human births (Randall and  Moos, 1993).
  
There is much weight on seasonal effects about  birth weights, for example in Germany (Miura and Richter, 1981), Israel (Chodick et al, 2007), in Iran (Khajavi et al., 2016) and in Chile (Torche and Corvalan, 2010). In the same spirit,  Jurges (2015) provides interesting results regarding (Ramadan) fasting effects on births on babies born in Muslim families. This report appears to be an interesting one since it is based on a very large database obtained from administrative sources and correlated to additional factors (such as day length). As the reportÕs title mention the author found almost no evidence of Ramadan effects on births. Subsequently the author suggests that other previous conclusions based Òon smaller samples from other countries must be interpreted with cautionÓ. We believe that, in our opinion, a number of issues need to be raised:

   \section{Data  }   \label{sec:dataanal}
   Consider the following points
   
\begin{itemize}
\item (i) The ÒsamplesÓ term is a little bit ambiguous since the paper to which one  is much referring  to, i.e. Almond and Mazumder (2011) is focused on whole populations (and not on a sample Ð which usually implies a selection process). A population could be smaller but this does not mean that the conclusions based on an exhaustive database could be biased such as it can be in a voluntarily selected sample.
\item (ii) Jurges fails to give credit to other recent (and very interesting) papers on this and highly related topics topic (Friger et al., 2009 or Herteliu et al., 2015); in the latter about 100 years, 35 429 days,  and  24 947 061 births were recorded and analyzed!
\item (iii) Since the Ordinary Least Squares (OLS) method was used, except for t-tests on regression parameters there is no other econometrical test (or vital information such as R2 regression analysis, models validity-Fisher test etc.) presented. Moreover there is no evidence about data statistical homogeneity, or about the distribution of variables used.

\item (iv) Depending on the distribution of assumed as continuous variables (e.g. birth weight) a semi-logarithmic approach could be a better solution instead of the presented-linear one. In the case of a non-linear approach the statistical significance of the covariates and the OLS assumption may have a significant impact on the practical results.
\item (v) Since the data used by Jurges (2015) study contains birthdays, there is a lack of precision induced by an over use the dichotomization (13 dummy variables!). Other papers took into consideration the overlap proportion of Ramadan (Almond and Mazumder, 2011) or a countdown approach (Herteliu et al., 2015) or a little bit more sophisticated models, as a cosinor (Friger et al., 2009, Cancho-Candela et al., 2007).
\end{itemize}

\section{Conclusions} 
In 
any seasonal adjustment filter, some cyclical variation will be misattributed to seasonal factors and vice versa.  Wright (2013) argues for using filters that constrain the seasonal factors to be more stable than the default filters and also for using filters that are based on estimation of a state-space model. Finally,  some evidence of predictability in revisions to seasonal factors is discussed

While the scientific sound of Jurges  paper and its topic maintain it to a high academic level, a part of the claimed conclusions could be a little bit inaccurate. We warn readers  
, authors, reviewers, and editors to take  Jurges (2015) conclusion with caution. In fact, in (Herteliu et al., 2015), noticeable effects dues to Lent and Nativity fast periods in which sexual activity is reprimanded by church leaders were demonstrated. Maybe, Muslim babies (in Germany) are different from Eastern Orthodox ones (in Romania)! A major question seems to be related to baby ÒproductionÓ: concerning ÒRamadanÓ per se, Friger et al. (2009) found a systematic increase in the number of births (200 009) during the Ramadan, in the Muslim population, - but not in the Jewish population in Israel. 

Thus, cultural constraints or psychological (Akuchekian et al., 2004) have to be taken during such analyses, exactly like in daily market index and company level stock return data (Bley and Saad, 2010). Notice, for completeness, that data analyzed by Roehner (2014) revealed a fall of about 15\% in suicide numbers during the month of Ramadan (with respect to same-non-Ramadan months). 
 Thus, to take into account baby deaths is another interesting question. 

Let our comment be also considered as a set of questions, beside a methodological one, raised by  recent contributions.
    \vskip0.5cm
 {\bf Acknowledgements}

 This paper is part of scientific activities in COST Action  TD1210 'Analyzing the
dynamics of information and knowledge landscapes'. This work by CH was co-financed by the
European Social Fund through the Sectorial Operational Programme Human Resources
Development 2007-2013, project number POSDRU/1.5/S/59184 Performance and excellence in
postdoctoral research in Romanian economics science domain

\end{document}